\documentclass[english]{article}
\usepackage[T1]{fontenc}
\usepackage[latin9]{inputenc}
\usepackage[a4paper]{geometry}
\usepackage{float}
\usepackage{amsmath}
\usepackage{amssymb}
\usepackage{graphicx}
\usepackage{cite}
\usepackage[unicode=true]{hyperref}
\usepackage{setspace}
\makeatletter

\floatstyle{ruled}
\newfloat{algorithm}{tbp}{loa}
\providecommand{\algorithmname}{Algorithm}
\floatname{algorithm}{\protect\algorithmname}
\makeatother

\begin{document}
\title{Delayed interactions in the noisy voter model through the periodic
polling mechanism}
\author{Aleksejus Kononovicius$^{1}$\thanks{corresponding author; email: \protect\href{mailto:aleksejus.kononovicius@tfai.vu.lt}{aleksejus.kononovicius@tfai.vu.lt};
website: \protect\href{https://kononovicius.lt}{kononovicius.lt}.}~, Rokas Astrauskas$^{2}$, Marijus Radavi\v{c}ius$^{3}$, Feliksas
Ivanauskas$^{2}$}
\date{{\small$^{1}$ - Institute of Theoretical Physics and Astronomy, Vilnius
University}\\
{\small$^{2}$ - Institute of Computer Science, Vilnius University}\\
{\small$^{3}$ - Institute of Applied Mathematics, Vilnius University}}

\maketitle
\begin{abstract}
We investigate the effects of delayed interactions on the stationary
distribution of the noisy voter model. We assume that the delayed
interactions occur through the periodic polling mechanism and replace
the original instantaneous two-agent interactions. In our analysis,
we require that the polling period aligns with the delay in announcing
poll outcomes. As expected, when the polling period is relatively
short, the model with delayed interactions is almost equivalent to
the original model. As the polling period increases, oscillatory behavior
emerges, but the model with delayed interactions still converges to
stationary distribution. The stationary distribution resembles a Beta-binomial
distribution, with its shape parameters scaling with the polling period.
The observed scaling behavior is non-monotonic. Namely, the shape
parameters peak at some intermediate polling period.
\end{abstract}

\section{Introduction}

In the physics realm, interactions among spatially distributed elements
are subject to temporal delay, as any physical interaction is inherently
bound by a finite propagation speed. Similarly, within the biological
sphere, communication between biological entities relies on biochemical
materials that also move at finite speeds \cite{Muller2003ABE,Dehghan2010CPC,Sargood2022BMB}.
Everyday social dynamics are equally affected by propagation constraints
arising from limited information processing capacities and finite
learning speeds \cite{Moreno2002JETh,Ahlin2007JDevEco,Kononovicius2019OB,Aghamolla2020JFinEco}.
The finite speed of information propagation across diverse systems
results in temporal delays, giving rise to intricate phenomena. These
delays manifest in phenomena such as stabilization of chaotic systems
\cite{Foss1996PRL,Yeung1999PRL,Pyragas2006PTRSA,Pati2023EPJB}, resonant
behavior in stochastic systems \cite{Inchiosa1997PRE,Ohira2000PRE,RouvasNicolis2007},
and pattern formation in evolutionary game dynamics and social systems
\cite{Ashcroft2013PRE,Touboul2019DCDsB,Dieci2022CSF,Roy2023SciRep},
among others. In real-world scenarios, public opinion polls often
take significant time to be conducted, processed, and subsequently
released to the public. Consequently, the polling mechanism can be
seen as a source of delays in the opinion formation process. Here,
we focus on the implications of information delays induced by the
periodic polling mechanism on the opinion formation process.

Modeling opinion formation is a primary concern within an emerging
subfield of statistical physics known as sociophysics \cite{Galam2008ModPhysC,Castellano2009RevModPhys,Abergel2017Springer,Jedrzejewski2019CRP,Redner2019CRP,Noorazar2020EPJP,Peralta2022}.
Opinion formation models describe the evolution of opinions within
artificially simulated societies as if they were describing magnetization
phenomena in spin systems. The voter model \cite{Clifford1973,Liggett1999}
stands out as one of the most thoroughly examined models in the field
of sociophysics. Introduced as a model for spatial conflict between
competing species, it has gained substantial popularity in opinion
dynamics and, for this reason, is known as the voter model \cite{Castellano2009RevModPhys}.
In the context of opinion dynamics, the spatial dimension from the
original model is replaced by a social network of individuals. Likewise,
the competing species from the original model are replaced by competing
opinions that the individuals could possess. From the statistical
physics perspective, we could interpret an individual as a kind of
social particle (referred as agents) and the distinct opinions as
the available states for the particles to be in. While multi-state
generalizations of the model exist \cite{Starnini2012JStat,Kononovicius2013EPL,Vazquez2019PRE},
most of the literature focuses on the other possible generalizations
of the voter model retaining the binary opinions \cite{Jedrzejewski2019CRP,Redner2019CRP,Noorazar2020EPJP}.

Here, we are particularly interested in a generalization known as
the noisy voter model \cite{Granovsky1995}. An analogous model was
introduced earlier in \cite{Kirman1993QJE}, hence this generalization
is occasionally referred to as Kirman's herding model. Both of these
approaches extend the voter model by allowing independent single-agent
transitions. In contrast to the voter model, the noisy voter model
doesn't converge to a fixed state (either full or partial consensus);
instead, it converges in a statistical sense to a broad stationary
distribution. Stationary distribution of the noisy voter model is
known to fit political party vote share distributions across various
elections quite well \cite{FernandezGracia2014PRL,Sano2016,Braha2017PlosOne,Kononovicius2017Complexity}.
Therefore, it can be seen as a minimal model for the political opinion
formation in the society. Consequently, the noisy voter model appears
to be a natural choice to explore the implications of information
delays induced by the periodic polling mechanism.

Latency in binary opinion formation processes, including the voter
model, was earlier considered in \cite{Lambiotte2009PRE}. Contrary
to our approach to temporal delays, Lambiotte~et~al. have considered
latency from an individual agent perspective. Namely, it was assumed
that individual agents become inactive immediately after changing
their state, but they may become activated again after some time.
In the latent opinion formation process, the inactive agents are effectively
equivalent to zealots, as they are unable to change their state, but
they may influence other agents. Later works have built upon the ideas
of the latent opinion formation process or from similar considerations
arrived at their independent approaches, but many of them working
towards studying physics-inspired aging and other state freezing effects
\cite{Artime2018PREaging,Peralta2020PhysA,Chen2020PRE,Latoski2022PRE}.
In our approach, latency creates an effect similar to zealotry \cite{Galam2007PhysA,Mobilia2007JStat,Kononovicius2014PhysA,Khalil2018PRE,Meyer2024NJP},
but with the difference that agents change their state without other
agents perceiving these changes until the announcement of the poll
outcome. Simulating polls was also addressed in a few earlier works
\cite{Levene2021IJF,Meyer2024NJP}, but these approaches were more
data-centric and therefore have not considered possible latency effects
or periodic driving of the electoral system.

This paper is structured as follows. In Section~\ref{sec:definition-micro-model},
we briefly discuss the original noisy voter model and then generalize
it by introducing the periodic polling mechanism. Having defined the
microscopic behavior rules, we introduce three distinct simulation
methods tailored for the generalized model with period polling, see
Section~\ref{sec:simulation-methods}. In Section~\ref{sec:stationary-distributions}
we explore the stationary poll outcome distributions both analytically
and numerically. In the short polling period limit, the delay has
a negligible effect. In the long polling period limit, the stationary
distribution of the generalized model is well approximated by the
Beta-binomial distribution with the shape parameters twice as large
as the independent transition rates. This finding suggests that the
periodic polling mechanism decreases the variance of the poll outcome
distribution. Yet the maximum of this effect is observed for some
intermediate polling period. Based on the approximation by a second-order
auto-regressive process \cite{Brockwell1991Springer}, we are able
to derive explicit analytical form of the scaling law as well as the
location of its maximum. In Section~\ref{sec:periodicity}, we analyze
periodic fluctuations induced by the periodic polling mechanism. While
the scaling behavior of the power spectral density follows a trivial
monotonic sigmoid-like functions, some interesting behavior is recovered
by examining the variance of the consecutive and the next-consecutive
poll swings. Finally, all findings are briefly summarized and future
outlook is given in Section~\ref{sec:conclusions}.

\section{Definition of the noisy voter model with the periodic polling mechanism\protect\label{sec:definition-micro-model}}

The noisy voter model describes the dynamics of a fixed number of
agents, denoted as $N$, switching between two possible states labeled
as ``$0$'' and ``$1$''. Agents switch their states independently
at a rate $\sigma_{i}$, where $i$ represents the label of the destination
state, or they imitate the states of their peers at a rate $h$. Since
only one agent changes its state at any given time, we can express
the system-wide transition rates with respect to the number of agents
in state ``$1$'', denoted by $X$, as follows:
\begin{equation}
\lambda\left(X\rightarrow X+1\right)=\lambda^{+}=\left(N-X\right)\left[\sigma_{1}+hX\right],\qquad\lambda\left(X\rightarrow X-1\right)=\lambda^{-}=X\left[\sigma_{0}+h\left(N-X\right)\right].
\end{equation}
Since the transition rates remain constant between the updates of
the system state $X$, simulating this model follows a standard approach
similar to any other homogeneous Poisson process. For example, this
model could be simulated by using one-step transition probability
approach \cite{VanKampen2007NorthHolland}, or by using Gillespie
method \cite{Anderson2007JCP}.

In the $N\rightarrow\infty$ limit it is trivial to show that $x=\frac{X}{N}$
is distributed according to the Beta distribution, $x\sim\mathcal{B}e\left(\frac{\sigma_{1}}{h},\frac{\sigma_{0}}{h}\right)$.
For the finite $N$, $X$ would be distributed according to the Beta-binomial
distribution, $X\sim\mathrm{BetaBin}\left(N,\frac{\sigma_{1}}{h},\frac{\sigma_{0}}{h}\right)$.
As the shape parameters of the stationary distribution depend only
on the ratio of $\sigma_{i}$ and $h$, we can simplify the model
by introducing dimensionless parameters $\varepsilon_{i}=\frac{\sigma_{i}}{h}$
and simulate the model in dimensionless time $t=ht^{\prime}$ (here
$t^{\prime}$ is the physical time measured in desired time units).

\begin{figure}[h]
\begin{centering}
\includegraphics[width=0.5\textwidth]{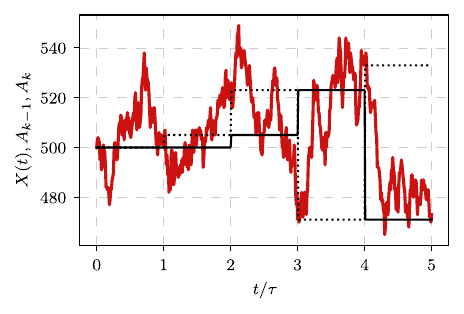}
\par\end{centering}
\caption{Dynamic interplay between the evolution of the system state $X\left(t\right)$
(red curve), the last announced poll outcome $A_{k-1}$ representing
the information about the system state available to agents (solid
black curve), and the latest poll outcome $A_{k}$ representing the
information to be revealed in the future (black dotted curve). Simulation
parameters were set as follows: $\varepsilon_{0}=\varepsilon_{1}=2$,
$\tau=5\cdot10^{-3}$, $N=10^{3}$, with initial conditions $A_{-1}=A_{0}=X\left(0\right)=500$.\protect\label{fig:explain-polls}}
\end{figure}

Let us generalize the noisy voter model by restricting imitative interactions
to occur solely through the periodic polls. We denote the polling
period as $\tau$. Let us assume that the polls perfectly reflect
the system state at the time of polling, but their outcomes are announced
with a delay. To keep the model simple, we assume that this delay
coincides with the polling period. Under these assumptions, the system-wide
transition rates become:
\begin{equation}
\lambda_{k}^{+}=\left(N-X\right)\left[\varepsilon_{1}+A_{k-1}\right],\qquad\lambda_{k}^{-}=X\left[\varepsilon_{0}+\left(N-A_{k-1}\right)\right],\label{eq:transition-rates}
\end{equation}
where $k=\left\lfloor \frac{t}{\tau}\right\rfloor $ is the index
of the last conducted poll, and $A_{k-1}$ is the last announced poll
outcome. In general $k$-th poll outcome would be defined as
\begin{equation}
A_{k}=X\left(\left\lfloor \frac{t}{\tau}\right\rfloor \tau\right).
\end{equation}
As implied by the form of the rates (\ref{eq:transition-rates}),
at time $t$, the most recently conducted poll outcome $A_{k}$ has
not yet been announced. Instead, the agents are aware of the outcome
of an earlier poll $A_{k-1}$, which we refer to as the last announced
poll outcome. For example, at $t=0$ the outcome $A_{-1}$ is announced
(it must be specified as a part of the initial condition), and the
outcome $A_{0}$ is recorded. Effectively it is also given as a part
of the initial condition, as $A_{0}=X\left(0\right)$. The outcome
of initial poll $A_{0}$ will be announced at $t=\tau$. Fig.~\ref{fig:explain-polls}
depicts a sample time series generated by the model, extending up
to $t=5\tau$. The red curve traces the evolution of the system state,
$X\left(t\right)$, while the black curves depict the last announced
poll outcome $A_{k-1}$ (solid curve) and the last conducted poll
outcome $A_{k}$ (dotted curve). At the start of each polling period,
at $t=k\tau$, the dotted curve intersects both the solid curve and
the red curve. As information about the last conducted poll is made
known, the solid curve catches up to the dotted curve. Immediately
afterwards, a new poll is conducted, which is represented by the dotted
curve catching up to the red curve. Between the subsequent polls,
the red curve exhibits fluctuations, predominantly converging towards
the solid black curve, reflecting incorporation of the available polling
information into the current system state.

Notably, upon closer examination of Fig.~\ref{fig:explain-polls},
there are indications of periodic oscillations arising due to the
periodic polling mechanism, even though the initial condition, $A_{-1}=A_{0}=X\left(0\right)$,
initially suppresses them. We will explore this effect in a more detail
in a subsequent section.

\section{Simulation methods\protect\label{sec:simulation-methods}}

Model driven by the rates (\ref{eq:transition-rates}) could be simulated
using one-step transition probability approach \cite{VanKampen2007NorthHolland},
with the condition that the time step is smaller than $\tau$ and
the ratio between $\tau$ and the time step is an integer. The issue
with this approach in the general case is that it is slow and it generates
biased samples \cite{Anderson2007JCP}. Gillespie method \cite{Anderson2007JCP}
could be employed as an approximation, but it would inaccurately represent
one transition per every polling period, specifically the transition
during which the crossover to the next polling period occurs. While
the potential error is likely negligible, as misrepresentation becomes
more noticeable only for small values of $\tau$, but the delay effect
induced by the polling mechanism is also smaller for smaller $\tau$
as well. Typically, for systems with delays a modified next reaction
method is used \cite{Anderson2007JCP}. In our case, this method has
performed approximately $4$ times slower than the Gillespie method,
but still, it has an advantage over the Gillespie method as it produces
time series without misrepresenting any transition. In this section,
we will discuss our adaptation of the Gillespie method for systems
with delays, as well as introduce a macroscopic simulation method
developed specifically for this model. We will also briefly touch
upon capturing the model dynamics using a one-dimensional Markov chain.

\subsection{Adapted Gillespie method for periodic polling with announcement delays}

We propose the adapted Gillespie method by combining the best features
of the Gillespie method and the next reaction method. Our adaption,
outline given in Algorithm~\ref{alg:adapted-gillespie}, is based
on the Gillespie method, but introduces delay $\tau$, the poll index
$k$, and the $k$-th poll outcome $A_{k}$. In Step~\ref{enu:gillespie-while-loop}
of the algorithm, the delay mechanism is introduced by building on
the idea of the internal reaction clock $R$ from the next reaction
method. This allows recalculation of the transition rates according
to updated recent poll outcomes. The conditional statement in Step~\ref{enu:gillespie-while-loop}
of the algorithm checks if a poll should be conducted before the next
transition (reaction, in the language of the original next reaction
method). The while loop is used to handle an edge case when more than
a single poll falls between $2$ transitions. This edge case arises
often when $\tau\lesssim\frac{1}{N^{2}}$.

\begin{algorithm}[h]
\caption{Adapted Gillespie method\protect\label{alg:adapted-gillespie}}

\begin{enumerate}
\item Set parameter values $\varepsilon_{0}$, $\varepsilon_{1}$, $N$,
$\tau$. Set desired initial conditions $A_{-1}$, $X\left(0\right)$.
Set the clock $t=0$. Set the current polling period index $k=0$.
Conduct the initial poll, $A_{0}=X\left(0\right)$.
\item Calculate the system-wide transition rates $\lambda^{+}$ and $\lambda^{-}$
according to Eq.~(\ref{eq:transition-rates}).
\item Calculate total transition rate $\lambda^{T}=\lambda^{+}+\lambda^{-}$.
\item Sample the time until the next reaction from an exponential distribution,
$\Delta t\sim\mathrm{Exp}\left(\lambda^{T}\right)$.\label{enu:gillespie-loop}
\item While $t+\Delta t\geq\left(k+1\right)\tau$:\label{enu:gillespie-while-loop}
\begin{itemize}
\item Increment the polling period index $k\rightarrow k+1$.
\item Conduct the $k$-th poll, $A_{k}=X\left(t\right)$.
\item Calculate the remaining time until the next reaction (according to
the internal reaction clock) $R=\lambda^{T}\left[t+\Delta t-k\tau\right]$.
\item Update $\lambda^{+}$, $\lambda^{-}$ according to Eq.~(\ref{eq:transition-rates}).
Update $\lambda^{T}$ accordingly.
\item Adjust the time until the next reaction $\Delta t=\frac{R}{\lambda^{T}}$.
\item Update the clock $t\rightarrow k\tau$
\end{itemize}
\item Update the clock $t\rightarrow t+\Delta t$.
\item Sample uniformly distributed random value $r\sim\mathcal{U}\left(0,\lambda^{T}\right)$.
If $r\leq\lambda^{+}$, set $X\left(t\right)=X\left(t-\Delta t\right)+1$.
Otherwise set $X\left(t\right)=X\left(t-\Delta t\right)-1$.
\item Go back to Step~\ref{enu:gillespie-loop} or end the simulation.
\end{enumerate}
\end{algorithm}

Python implementation of this method for the noisy voter model with
delayed interactions is available at \cite{DelayGithub}.

\subsection{Macroscopic simulation method\protect\label{subsec:macro-method}}

This simulation method relies on an observation that the imitation
term $A_{k-1}$ remains constant throughout the polling interval.
This enables us to introduce the effective individual agent transition
rates that remain constant for the duration of the $k$-th polling
period:
\begin{equation}
\varepsilon_{1}^{\left(k\right)}=\varepsilon_{1}+A_{k-1},\quad\varepsilon_{0}^{\left(k\right)}=\varepsilon_{0}+\left(N-A_{k-1}\right).\label{eq:individual-effective-rates}
\end{equation}
These effective rates encompass both the truly independent transitions
and the imitative behavior induced by the knowledge of the last announced
poll outcome. This effect is somewhat reminiscent of the peer pressure
exerted by zealots \cite{Galam2007PhysA,Mobilia2007JStat,Kononovicius2014PhysA,Khalil2018PRE,Meyer2024NJP},
although in our case the agents themselves still change their state,
only their knowledge about the other agents remains conserved for
the duration of the polling period. Consequently, the system-wide
transition rates during the polling period $k$ would be given by
\begin{equation}
\lambda_{k}^{+}=\left(N-X\right)\varepsilon_{1}^{\left(k\right)},\quad\lambda_{k}^{-}=X\varepsilon_{0}^{\left(k\right)}.\label{eq:system-wide-effective-rates}
\end{equation}
The form of the system-wide transition rates suggests that each agent
operates independently of others at all times, as they consider the
available polling information. Upon the announcement of a new poll
outcome, the transition rates get updated. Hence, we can approach
the analysis of this model from the standpoint of an individual agent,
and concentrating only on the current polling period.

In examining the behavior of a single agent, and given that the agent
can occupy one of two possible states, the dynamics can be analyzed
as a two-state Markov chain. Given the effective individual agent
transition rates, Eq.~(\ref{eq:individual-effective-rates}), we
can formulate the corresponding left stochastic transition matrix
governing the transitions of an individual agent over an infinitesimally
short time interval $\Delta t$:
\begin{equation}
\boldsymbol{Q}=\left(\begin{array}{cc}
1-\varepsilon_{0}^{\left(k\right)}\Delta t & \varepsilon_{1}^{\left(k\right)}\Delta t\\
\varepsilon_{0}^{\left(k\right)}\Delta t & 1-\varepsilon_{1}^{\left(k\right)}\Delta t
\end{array}\right).
\end{equation}
By solving the eigenproblem with respect to $\boldsymbol{Q}$, we
can infer that the probability to observe an agent in state ``$1$''
after $m$ steps is given by
\begin{equation}
P_{1}\left(m|P_{1}\left(0\right)\right)=\frac{\varepsilon_{1}^{\left(k\right)}}{\varepsilon_{0}^{\left(k\right)}+\varepsilon_{1}^{\left(k\right)}}+\frac{\varepsilon_{0}^{\left(k\right)}P_{1}\left(0\right)-\varepsilon_{1}^{\left(k\right)}\left[1-P_{1}\left(0\right)\right]}{\varepsilon_{0}^{\left(k\right)}+\varepsilon_{1}^{\left(k\right)}}\left[1-\left(\varepsilon_{0}^{\left(k\right)}+\varepsilon_{1}^{\left(k\right)}\right)\Delta t\right]^{m}.
\end{equation}
In the above $P_{1}\left(0\right)$ represents the ``initial'' condition
of the Markov chain describing individual agent dynamics. Typically,
$P_{1}\left(0\right)$ assumes a value of $1$ if the agent under
consideration is initially in the ``$1$'' state, or $0$ otherwise.
Additionally, it proves convenient to introduce notation $P_{1}\left(\infty\right)$
which denotes the stationary probability of observing an agent in
the ``$1$'' state,
\begin{equation}
P_{1}\left(\infty\right)=\frac{\varepsilon_{1}^{\left(k\right)}}{\varepsilon_{0}^{\left(k\right)}+\varepsilon_{1}^{\left(k\right)}}=\frac{\varepsilon_{1}+A_{k-1}}{\varepsilon_{0}+\varepsilon_{1}+N}.\label{eq:single-agent-prob-stationary}
\end{equation}

By taking the continuous time limit, i.e., letting $\Delta t\rightarrow0$
and $m\rightarrow\infty$ (with $s=m\Delta t=\mathrm{const}$), we
obtain the conditional probability to observe an agent in the ``$1$''
state after time span $s$,
\begin{equation}
P_{1}\left(s|P_{1}\left(0\right)\right)=P_{1}\left(\infty\right)+\left[P_{1}\left(0\right)-P_{1}\left(\infty\right)\right]\exp\left[-\left(\varepsilon_{0}+\varepsilon_{1}+N\right)s\right].\label{eq:single-agent-prob}
\end{equation}

We can use Eq.~(\ref{eq:single-agent-prob}) to simulate the behavior
of all $N$ agents without resorting to the time-consuming direct
simulation of the noisy voter model with periodic polling mechanism.
Let $X\left(t\right)$ denote the system state at some arbitrary time
$t$, and let $s$ be a positive time increment such that $k\tau\leq t<t+s\leq\left(k+1\right)\tau$.
Then, $X\left(t+s\right)$ can be sampled by adding two binomial random
variables
\begin{equation}
X\left(t+s\right)=B_{1\rightarrow1}\left[X\left(t\right),P_{1}\left(s|1\right)\right]+B_{0\rightarrow1}\left[N-X\left(t\right),P_{1}\left(s|0\right)\right].\label{eq:main-macro-rel}
\end{equation}
In the expression above, $B_{1\rightarrow1}\left[\ldots\right]$ corresponds
to the count of agents that were in state ``$1$'' at time $t$
and ended up in state ``$1$'' at time $t+s$. These agents may
have remained in state ``$1$'' for the duration $s$, or they might
have exited and subsequently returned to state ``$1$''. In this
setup, the specific evolution of an individual agent's state doesn't
influence the outcome; only the initial and final states matter. Given
there were $X\left(t\right)$ agents in state ``$1$'' at time $t$,
and the probability that an agent starting in state ``$1$'' will
end up in state ``$1$'' is given by $P_{1}\left(s|1\right)$, then
$B_{1\rightarrow1}\left[\ldots\right]$ is an outcome of $X\left(t\right)$
Bernoulli trials with a success probability of $P_{1}\left(s|1\right)$.
Similarly, $B_{0\rightarrow1}\left[\ldots\right]$ is an outcome of
$N-X\left(t\right)$ Bernoulli trials with a success probability of
$P_{1}\left(s|0\right)$.

This approach is most efficient when $t=k\tau$ and $s=\tau$, although
finer-scale simulations are also possible for $s<\tau$. As long as
the sampling period $s$ encompasses a large number of transitions,
this method proves to be more efficient than a direct simulation without
compromising quality of the sampled time series. The detailed outline
of the macroscopic simulation method is provided in Algorithm~\ref{alg:macro-algorithm}.

\begin{algorithm}[h]
\caption{Macroscopic simulation method\protect\label{alg:macro-algorithm}}

\begin{enumerate}
\item Set parameter values $\varepsilon_{0}$, $\varepsilon_{1}$, $N$,
$\tau$. Set desired initial conditions $A_{-1}$, $X\left(0\right)$.
Set desired sampling period $s$ (note that $\tau/s$ must be a positive
integer). Set the clock $t=0$. Set the current polling period index
$k=0$.
\item Calculate the effective transition rates $\varepsilon_{0}^{\left(k\right)}=\varepsilon_{0}+\left(N-A_{k-1}\right)$,
$\varepsilon_{1}^{\left(k\right)}=\varepsilon_{1}+A_{k-1}$.\label{enu:binom-start-loop}
\item Calculate the transition probabilities $P_{1}\left(s|1\right)$ and
$P_{1}\left(s|0\right)$.
\item Conduct the $k$-th poll, $A_{k}=X\left(t\right)$.
\item Sample two binomial random values $B_{1\rightarrow1}\sim\mathrm{Binom}\left[X\left(t\right),P_{1}\left(s|1\right)\right]$
and $B_{0\rightarrow1}\sim\mathrm{Binom}\left[N-X\left(t\right),P_{1}\left(s|0\right)\right]$.\label{enu:internal-loop}
\item Update the system state $X\left(t+s\right)=B_{1\rightarrow1}+B_{0\rightarrow1}$.
\item Update the clock $t\rightarrow t+s$.
\item If $t<\left(k+1\right)\tau$, go back to Step~\ref{enu:internal-loop}.
\item Increment the polling period index $k\rightarrow k+1$.
\item Go back to Step~\ref{enu:binom-start-loop} or end the simulation.
\end{enumerate}
\end{algorithm}

Python implementation of this method for the noisy voter model with
delayed interactions is available at \cite{DelayGithub}.

\subsection{Comparison of the Monte-Carlo simulation methods}

Both of the methods discussed earlier are Monte Carlo simulation methods.
In order to obtain the temporal dependence of statistical moments
or the stationary distribution, it is necessary to conduct multitude
simulations using the same parameter set and subsequently average
over the ensemble. Comparing the results obtained from simulations
using these methods allows us to verify the validity of the macroscopic
simulation method, which may not be immediately evident.

\begin{figure}
\begin{centering}
\includegraphics[width=0.9\textwidth]{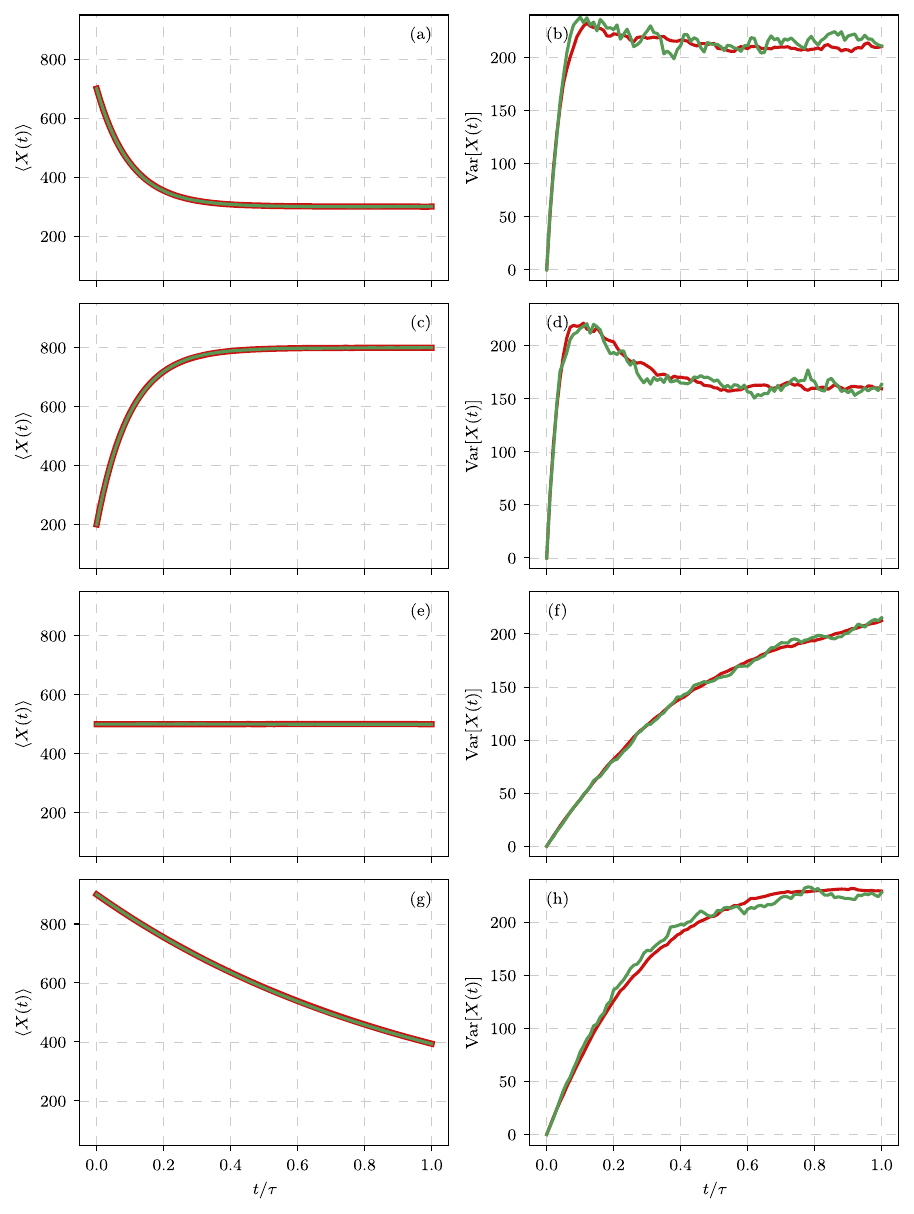}
\par\end{centering}
\caption{The evolution of statistical moments, mean ((a), (c), (e) and (g))
and variance ((b), (d), (f) and (h)), of numerically simulated ensembles.
Different curves correspond to results obtained from the two simulation
methods: red curve corresponds to the macroscopic simulation method
(with ensemble size of $10^{4}$), green curve corresponds to the
adapted Gillespie method (with ensemble size of $10^{3}$). Different
pairs of plots were obtained with the different parameter sets: $\varepsilon_{0}=\varepsilon_{1}=0.5$,
$\tau=10^{-2}$, $A_{-1}=300$ and $A_{0}=X\left(0\right)=700$ ((a)
and (b)); $\varepsilon_{0}=\varepsilon_{1}=2$, $\tau=10^{-2}$, $A_{-1}=800$
and $A_{0}=X\left(0\right)=200$ ((c) and (d)); $\varepsilon_{0}=0.5$,
$\varepsilon_{1}=0.5$, $\tau=10^{-3}$ and $A_{-1}=A_{0}=X\left(0\right)=500$
((e) and (f)); $\varepsilon_{0}=\varepsilon_{1}=2$, $\tau=10^{-3}$,
$A_{-1}=100$ and $A_{0}=X\left(0\right)=900$ ((g) and (h)). Shared
parameter values: $N=10^{3}$.\protect\label{fig:test-approx}}
\end{figure}

In the different simulations shown in Fig.~\ref{fig:test-approx}
we keep $N$ fixed and equal to $10^{3}$. We systematically vary
the values of $\varepsilon_{i}$ and $\tau$ parameters. While the
initial conditions are purposefully selected to be very different
in order to emphasize their importance on the values of mean and variance
reached during the polling period. For all distinct cases the results
of numerical simulations using both methods match reasonably well.
So well that we are forced to make the red $\left\langle X\left(t\right)\right\rangle $
curve (obtained using the macroscopic simulation method) thicker.
Fig.~\ref{fig:test-approx}~(a) and (b) show how the mean and variance
evolve for the base parameter set. The selected value of $\tau=10^{-2}$
appears to be sufficient for the statistical moments to converge towards
their stationary values; the mean approaches $A_{-1}$. As the delay
$\tau$ is kept the same in Fig.~\ref{fig:test-approx}~(c) and
(d), the statistical moments still converge to their respective stationary
values. In Fig.~\ref{fig:test-approx}~(d) we can clearly observe
localization phenomenon as the ensemble variance temporarily increases
before converging to the stationary value. From Fig.~\ref{fig:test-approx}~(e)-(h)
it is evident that for shorter delay, $\tau=10^{-3}$, the statistical
moments fail to converge their respective stationary values: instead
some intermediate values are reached. From Fig.~\ref{fig:test-approx}
it is not clear what impact $\varepsilon_{i}$ parameters have, while
the initial conditions appear to be extremely important. This was
expected as the macroscopic simulation method takes the effective
rates as its input.

Producing Fig.~\ref{fig:test-approx} allows us to at least approximately
compare the speed of the methods. It took couple of seconds to obtain
all of the results using the macroscopic simulation method, while
it took couple of minutes using the adapted Gillespie method. Given
difference in the ensemble sizes, macroscopic simulation method produces
the results roughly $10^{2}$ times faster for the considered parameter
sets and the selected time resolution. The difference in favor of
the macroscopic simulation method was expected as it doesn't simulate
individual transitions, only $X\left(t\right)$ values for desired
$t$.

In the subsequent sections, we present the results obtained by simulating
large a number of polling periods. Wherever feasible, we will use
both simulation methods to reinforce validity of the obtained results
as well as to show the equivalence of both Monte-Carlo simulation
methods.

\subsection{Semi-analytical approach based on the transition matrix for poll
outcomes\protect\label{subsec:semi-analytical-approach}}

If we focus on the poll outcomes $A_{i}$, the model can be treated
as a second-order Markov chain, as the distribution of $A_{i}$ is
conditioned on both $A_{i-1}$ and $A_{i-2}$. Note thate finite $N$
implies that the phase space of the model is also finite. This allows
us to reduce the model into a first-order Markov chain instead of
the second-order Markov chain. Let us proceed to derive an expression
for the left stochastic transition matrix elements of the first-order
Markov chain.

Upon reducing the second-order Markov chain, we effectively introduce
two-dimensional system state $\left(A_{i},A_{i-1}\right)$. As $A_{i}\in\left[0,N\right]$,
we can uniquely map the two-dimensional system state into one-dimensional
index $K_{i}$:
\begin{equation}
K_{i}=1+A_{i}+\left(N+1\right)\cdot A_{i-1}.
\end{equation}
Index $K_{i}$ corresponds to the row or column indices of the transition
matrix $\boldsymbol{T}$. Given that $A_{i}\in\left[0,N\right]$,
we have that $K_{i}\in\left[1,\left(N+1\right)^{2}\right]$. This
implies that the transition matrix will have $\left(N+1\right)^{4}$
elements, although only $\left(N+1\right)^{2}$ of them will be non-zero.
The one-dimensional index $K_{i}$ also uniquely maps to the two-dimensional
system state $\left(A_{i},A_{i-1}\right)$:
\begin{equation}
A_{i}=K_{i}-1-\left(N+1\right)\left\lfloor \frac{K_{i}-1}{N+1}\right\rfloor ,\qquad A_{i-1}=\left\lfloor \frac{K_{i}-1}{N+1}\right\rfloor .
\end{equation}
In the indexing scheme introduced above, the $\left(K,M\right)$ element
of the left stochastic transition matrix $\boldsymbol{T}$ representing
$M\rightarrow K$ transition is given by
\begin{align}
T_{K,M} & =P\left[M\rightarrow K\right]=\nonumber \\
 & =P\left[\left(M-1-\left(N+1\right)\left\lfloor \frac{M-1}{N+1}\right\rfloor ,\left\lfloor \frac{M-1}{N+1}\right\rfloor \right)\rightarrow\left(K-1-\left(N+1\right)\left\lfloor \frac{K-1}{N+1}\right\rfloor ,\left\lfloor \frac{K-1}{N+1}\right\rfloor \right)\right]=\nonumber \\
 & =\begin{cases}
P\left[K-1-\left(N+1\right)\left\lfloor \frac{K-1}{N+1}\right\rfloor |\left\lfloor \frac{K-1}{N+1}\right\rfloor ,\left\lfloor \frac{M-1}{N+1}\right\rfloor ,\tau\right] & \text{if }\left\lfloor \frac{K-1}{N+1}\right\rfloor =M-1-\left(N+1\right)\left\lfloor \frac{M-1}{N+1}\right\rfloor ,\\
0 & \text{otherwise.}
\end{cases}
\end{align}
The conditional probability in the above is given by
\begin{equation}
P\left[A_{i+1}|A_{i},A_{i-1},\tau\right]=\sum_{k=0}^{A_{i+1}}p_{\mathrm{Binom}}\left[k,A_{i},P_{1}\left(\tau|1,A_{i-1}\right)\right]\cdot p_{\mathrm{Binom}}\left[A_{i+1}-k,N-A_{i},P_{1}\left(\tau|0,A_{i-1}\right)\right].
\end{equation}
In the above, $p_{\mathrm{Binom}}\left(k,N,p\right)$ represents the
probability mass function of the Binomial distribution with $N$ trials
and success probability $p$, while $P_{1}\left(\ldots\right)$ corresponds
to Eq.~(\ref{eq:single-agent-prob}) additionally conditioned that
the last announced poll outcome was $A_{i-1}$. The last announced
poll outcome is not explicitly present in Eq.~(\ref{eq:single-agent-prob});
however, it is implicitly present as a part of $P_{1}\left(\infty\right)$
and $\varepsilon_{i}^{\left(k\right)}$.

This approach provides an alternative semi-analytical method for simulating
the model. The primary drawback of this method is that it is very
time consuming, making it feasible only for small $N$. However, solving
the eigenproblem with respect to $\boldsymbol{T}$ allows obtaining
the exact stationary distribution or the entire temporal evolution
of the distribution for the selected parameters. The methods discussed
earlier are faster, but they do not yield exact results.

Python implementation of this approach for the noisy voter model with
delayed interactions is available at \cite{DelayGithub}.

\section{Stationary poll outcome distributions\protect\label{sec:stationary-distributions}}

As discussed in the previous section, the outcome of the next poll
$A_{k+1}$ for an arbitrary polling interval $\tau$ depends on the
last announced poll outcome $A_{k-1}$ and the system state at the
start of the polling period $X\left(k\tau\right)$, which corresponds
to $A_{k}$. Namely, the model behaves as a second-order Markov chain,
yet it can be reduced to the first-order Markov chain for finite $N$.
Determining eigenvectors and eigenvalues of the associated transition
matrix yields the complete information about the evolution of the
poll outcome distribution and also the exact stationary poll outcome
distribution. However, an analytical solution of the eigenproblem
is elusive, necessitating a numerical approach. The numerical solution
of the eigenproblem is somewhat time-consuming and is only practical
for small $N$. Alternatively, we can explore other approaches to
derive analytical approximations for stationary poll outcome distributions.
This objective drives the focus of this section.

\begin{figure}[h]
\begin{centering}
\includegraphics[width=0.5\textwidth]{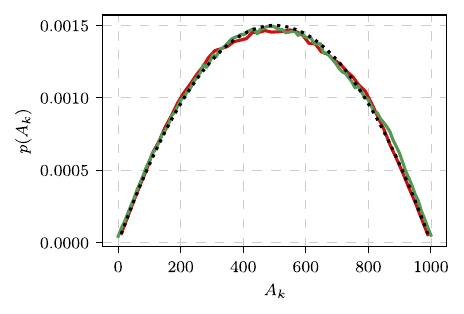}
\par\end{centering}
\caption{Stationary poll outcome distribution of the model with small $\tau$.
Numerical simulations were conducted using the macroscopic simulation
method (red curve) and the adapted Gillespie method (green curve).
Black dotted curve shows the probability mass function of the $\mathrm{BetaBin\left(N,\varepsilon_{1},\varepsilon_{0}\right)}$.
Simulation parameters: $\varepsilon_{0}=\varepsilon_{1}=2$, $\tau=10^{-7}$,
$N=10^{3}$.\protect\label{fig:small-tau-dist}}
\end{figure}

From Eq.~(\ref{eq:main-macro-rel}), it can be shown that the conditional
mean of $A_{k+1}$ with respect to $A_{k}$ and $A_{k-1}$ is given
by
\begin{equation}
\left\langle A_{k+1}|A_{k},A_{k-1}\right\rangle =\varphi_{1}A_{k}+\left(1-\varphi_{1}\right)\varphi_{2}\left(\varepsilon_{1}+A_{k-1}\right)\label{eq:cond-mean}
\end{equation}
with $\varphi_{1}=\exp\left[-\left(\varepsilon_{0}+\varepsilon_{1}+N\right)\tau\right]$
and $\varphi_{2}=\frac{N}{\varepsilon_{0}+\varepsilon_{1}+N}$. Observe
that if $\tau\rightarrow0$, then $\varphi_{1}\rightarrow1$ and the
conditional mean becomes independent of $A_{k-1}$. For small values
of $\tau$, the impact of the information delay should be minimal,
as the poll outcomes are updated nearly as frequently as the system
state. This suggests that the delay would have a negligible effect
on the poll outcomes. Consequently, when no or few transitions occur
during a single polling period, which is often the case with $\tau\ll\frac{2}{N\left(\varepsilon_{0}+\varepsilon_{1}+N\right)}$,
the model with delayed interactions should be almost equivalent to
the noisy voter model. The equivalence implies that the stationary
poll outcome distribution should be the same as the stationary distribution
the noisy voter model would have, i.e., $A_{\infty}\sim\mathrm{BetaBin}\left(N,\varepsilon_{1},\varepsilon_{0}\right)$.
This intuition is confirmed by numerical simulation (see Fig.~\ref{fig:small-tau-dist}).

\begin{figure}[th]
\begin{centering}
\includegraphics[width=0.9\textwidth]{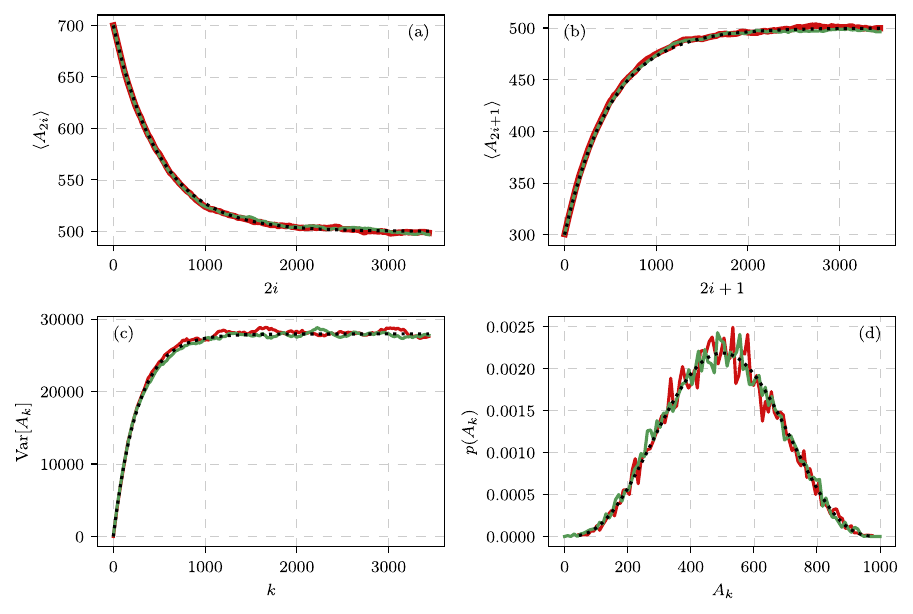}
\par\end{centering}
\caption{Evolution of the statistical moments and the stationary distribution
of the model with large $\tau$. (a) and (b) depict the evolution
of the mean for even and odd poll indices, respectively, while (c)
shows the evolution of the variance. (d) shows the stationary distribution.
The red curve represents simulation results obtained using the macroscopic
simulation method, while the green curve - results from the adapted
Gillespie method. Black dotted curves correspond to analytical predictions:
Eq.~(\ref{eq:big-tau-even-mean}) for (a), Eq.~(\ref{eq:big-tau-odd-mean})
for (b), Eq.~(\ref{eq:big-tau-even-var}) for (c), and the probability
mass function of $\mathrm{BetaBin}\left(N,2\varepsilon_{1},2\varepsilon_{0}\right)$
for (d). Simulation parameters: $\varepsilon_{0}=\varepsilon_{1}=2$,
$N=10^{3}$, $\tau=0.03$, $A_{-1}=300$, and $A_{0}=X\left(0\right)=700$.\protect\label{fig:big-tau-dist}}
\end{figure}

Likewise, in the large $\tau$ limit, i.e., $\tau\rightarrow\infty$,
we have that $\varphi_{1}\rightarrow0$. Therefore, in this limit
the conditional mean becomes independent of $A_{k}$. Intuitively,
this suggests that in this limit, the model with delayed interactions
behaves like two nearly independent Markov chains. One chain corresponds
to the even poll indices, and the other to the odd poll indices. These
Markov chains would be essentially identical in all aspects except
for their initial conditions. It can be shown (see Appendix~\ref{sec:large-tau-limit})
that the stationary mean in the large $\tau$ limit is given by
\begin{equation}
\left\langle A_{\infty}\right\rangle =\frac{N\varepsilon_{1}}{\varepsilon_{0}+\varepsilon_{1}}.\label{eq:big-tau-stationary-mean}
\end{equation}
For the large number of agents $N\gg\left(\varepsilon_{0}+\varepsilon_{1}\right)$,
the stationary variance in the large $\tau$ limit can be approximated
by
\begin{equation}
\mathrm{Var}\left[A_{\infty}\right]\approx\frac{N\varepsilon_{1}\varepsilon_{0}\left(2\varepsilon_{0}+2\varepsilon_{1}+N\right)}{\left(\varepsilon_{0}+\varepsilon_{1}\right)^{2}\left(2\varepsilon_{0}+2\varepsilon_{1}+1\right)}.\label{eq:big-tau-approx-stationary-var}
\end{equation}
The expressions for the stationary moments suggest that the stationary
poll outcome distribution in the large $\tau$ limit can be well approximated
by $\mathrm{BetaBin}\left(N,2\varepsilon_{1},2\varepsilon_{0}\right)$
distribution. In Appendix~\ref{sec:large-tau-limit} we have derived
not only the exact stationary moments but have also determined their
temporal evolution, i.e., we have obtained $\left\langle A_{k}\right\rangle $
and $\mathrm{Var}\left[A_{k}\right]$ expressions with arbitrary $k$.
In Fig.~\ref{fig:big-tau-dist} we show that the obtained analytical
expressions align with numerical simulation results rather well.

The results for the small $\tau$ and large $\tau$ limits prompt
us to posit that the stationary distribution of the model with delayed
interactions is a Beta-binomial for all possible $\tau$. Only the
shape parameters of the distribution change with $\tau$ according
to some scaling law.

\begin{figure}[h]
\begin{centering}
\includegraphics[width=0.5\textwidth]{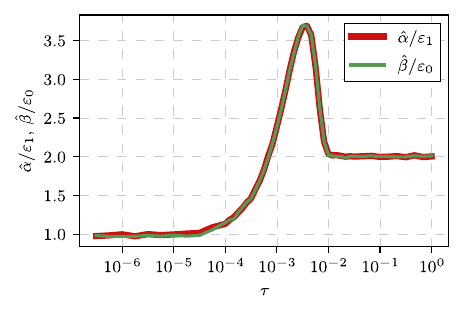}
\par\end{centering}
\caption{Scaling behavior of the normalized shape parameter estimates with
respect to the polling period. Simulation parameters: $\varepsilon_{0}=2$,
$\varepsilon_{1}=0.5$, $N=10^{3}$.\protect\label{fig:alpha-beta-epsi-asym}}
\end{figure}

Let us average Eq.~(\ref{eq:cond-mean}) over stationary distribution,
\begin{equation}
\left\langle A_{\infty}\right\rangle =\varphi_{1}\left\langle A_{\infty}\right\rangle +\left(1-\varphi_{1}\right)\varphi_{2}\left(\varepsilon_{1}+\left\langle A_{\infty}\right\rangle \right).
\end{equation}
Solving the above with respect to $\left\langle A_{\infty}\right\rangle $,
yields
\begin{equation}
\left\langle A_{\infty}\right\rangle =\frac{\varphi_{2}\varepsilon_{1}}{1-\varphi_{2}}=\frac{N\varepsilon_{1}}{\varepsilon_{0}+\varepsilon_{1}}.
\end{equation}
Which is identical to the stationary mean obtained for the small $\tau$
and large $\tau$ limits. The fact that $\left\langle A_{\infty}\right\rangle $
does not depend on $\tau$ indicates that both shape parameters of
the stationary distribution follow the same scaling law $L\left(\tau\right)$.
In other words, we have that 
\begin{equation}
\hat{\alpha}\left(\tau\right)=\varepsilon_{1}\cdot L\left(\tau\right)\quad\text{and}\quad\hat{\beta}\left(\tau\right)=\varepsilon_{0}\cdot L\left(\tau\right),
\end{equation}
here $\hat{\alpha}$ and $\hat{\beta}$ denote best estimates of the
shape parameters of the Beta-binomial distribution. If $\hat{\alpha}$
and $\hat{\beta}$ would follow different scaling laws, $\left\langle A_{\infty}\right\rangle $
would depend on $\tau$. In Fig.~\ref{fig:alpha-beta-epsi-asym},
we observe that $\hat{\alpha}/\varepsilon_{1}$ and $\hat{\beta}/\varepsilon_{0}$
obtained by numerical simulation align well, thus supporting the idea
of the shared scaling law.

To facilitate further analysis, the goal of which is to determine
$L\left(\tau\right)$, let us consider poll outcomes centered on the
stationary mean,
\begin{equation}
\tilde{A}_{k}=A_{k}-\left\langle A_{\infty}\right\rangle .
\end{equation}
Rewriting Eq.~(\ref{eq:cond-mean}) with respect to $\tilde{A}_{k}$
yields
\begin{equation}
\left\langle \tilde{A}_{k+1}|\tilde{A}_{k},\tilde{A}_{k-1}\right\rangle =\varphi_{1}\tilde{A}_{k}+\left(1-\varphi_{1}\right)\varphi_{2}\tilde{A}_{k-1}.\label{eq:ar2-centered-mean}
\end{equation}
From Eq.~(\ref{eq:main-macro-rel}), it follows that any deviations
$\xi_{k}$ of $\tilde{A}_{k}$ from their conditional expected values
$\left\langle \tilde{A}_{k}|\tilde{A}_{k-1},\tilde{A}_{k-2}\right\rangle $
are conditionally independent and hence uncorrelated. We could assume
that $\xi_{k}$ follows a normal distribution with zero mean and stationary
variance $\sigma_{\xi}^{2}$. Under this assumption, Eq.~(\ref{eq:ar2-centered-mean})
could be understood as a normal approximation to the macroscopic simulation
method. Yet for further derivation this assumption is not strictly
necessary. Let us then approximate $\tilde{A}_{k}$ process by a stationary
second-order auto-regressive process \cite{Brockwell1991Springer}
of the following form:
\begin{equation}
\tilde{A}_{k+1}=\varphi_{1}\tilde{A}_{k}+\left(1-\varphi_{1}\right)\varphi_{2}\tilde{A}_{k-1}+\xi_{k+1}.
\end{equation}
From Yule-Walker equations \cite{Brockwell1991Springer}, we can determine
stationary correlation between $\tilde{A}_{k+1}$ and $\tilde{A}_{k}$
denoted by $\rho_{1}$ and stationary correlation between $\tilde{A}_{k+1}$
and $\tilde{A}_{k-1}$ denoted by $\rho_{2}$:
\begin{equation}
\rho_{1}=\frac{\varphi_{1}}{1-\left(1-\varphi_{1}\right)\varphi_{2}},\qquad\rho_{2}=\left(1-\varphi_{1}\right)\varphi_{2}+\frac{\varphi_{1}^{2}}{1-\left(1-\varphi_{1}\right)\varphi_{2}}.
\end{equation}
And in turn, the stationary variance of the poll outcome distribution
would be a solution of
\begin{equation}
\mathrm{Var}\left[A_{\infty}\right]\left(1-\varphi_{1}\rho_{1}-\left[1-\varphi_{1}\right]\varphi_{2}\rho_{2}\right)=\mathrm{Var}\left[\xi_{\infty}\right].\label{eq:variance-problem}
\end{equation}

To proceed further let us determine $\mathrm{Var}\left[\xi_{\infty}\right]$.
From Eq.~(\ref{eq:main-macro-rel}) with $s=\tau$, we have that
\begin{align}
\mathrm{Var}\left[\xi_{k}|\tilde{A}_{k},\tilde{A}_{k-1}\right] & =\mathrm{Var}\left[B_{1\rightarrow1}|\tilde{A}_{k},\tilde{A}_{k-1}\right]+\mathrm{Var}\left[B_{0\rightarrow1}|\tilde{A}_{k},\tilde{A}_{k-1}\right]=\nonumber \\
 & =\psi_{0}+\psi_{1}\tilde{A}_{k}+\psi_{2}\tilde{A}_{k-1}+\psi_{12}\tilde{A}_{k}\tilde{A}_{k-1}+\psi_{22}\tilde{A}_{k-1}^{2}.\label{eq:var-cond}
\end{align}
The coefficients above are given by
\begin{align}
\psi_{0} & =\frac{N\varepsilon_{0}\varepsilon_{1}\left(1-\varphi_{1}^{2}\right)}{\left(\varepsilon_{0}+\varepsilon_{1}\right)^{2}},\qquad\psi_{1}=\frac{\left(\varepsilon_{0}-\varepsilon_{1}\right)\varphi_{1}\left(1-\varphi_{1}\right)}{\varepsilon_{0}+\varepsilon_{1}},\qquad\psi_{2}=\frac{N\left(\varepsilon_{0}-\varepsilon_{1}\right)\left(1-\varphi_{1}\right)}{\left(\varepsilon_{0}+\varepsilon_{1}+N\right)\left(\varepsilon_{0}+\varepsilon_{1}\right)},\nonumber \\
\psi_{12} & =-\frac{2\varphi_{1}\left(1-\varphi_{1}\right)}{\varepsilon_{0}+\varepsilon_{1}+N},\qquad\psi_{22}=-\frac{N\left(1-\varphi_{1}\right)^{2}}{\left(\varepsilon_{0}+\varepsilon_{1}+N\right)^{2}}.\label{eq:psi-params}
\end{align}
Averaging Eq.~(\ref{eq:var-cond}) over the stationary distribution
yields
\begin{equation}
\mathrm{Var}\left[\xi_{\infty}\right]=\left\langle \mathrm{Var}\left[\xi_{k}|\tilde{A}_{k},\tilde{A}_{k-1}\right]\right\rangle =\psi_{0}+\left(\psi_{12}\rho_{1}+\psi_{22}\right)\mathrm{Var}\left[A_{\infty}\right].\label{eq:variance-ar2-deviations}
\end{equation}

Inserting Eq.~(\ref{eq:variance-ar2-deviations}) into Eq.~(\ref{eq:variance-problem})
and solving it with respect to $\mathrm{Var}\left[A_{\infty}\right]$
yields
\begin{equation}
\mathrm{Var}\left[A_{\infty}\right]=\frac{\psi_{0}}{1-\left(\varphi_{1}+\psi_{12}\right)\rho_{1}-\left(1-\varphi_{1}\right)\varphi_{2}\rho_{2}-\psi_{22}}.\label{eq:variance-ar2-stationary}
\end{equation}
Taking $\tau\rightarrow0$ and $\tau\rightarrow\infty$ limits of
the above yields the expected results. For $\tau\rightarrow\infty$,
Eq.~\ref{eq:big-tau-stationary-var-exact} is recovered. Taking $\tau\rightarrow0$
limit yields the expression for variance of the $\mathrm{BetaBin}\left(N,\varepsilon_{1},\varepsilon_{0}\right)$.

\begin{figure}[th]
\begin{centering}
\includegraphics[width=0.9\textwidth]{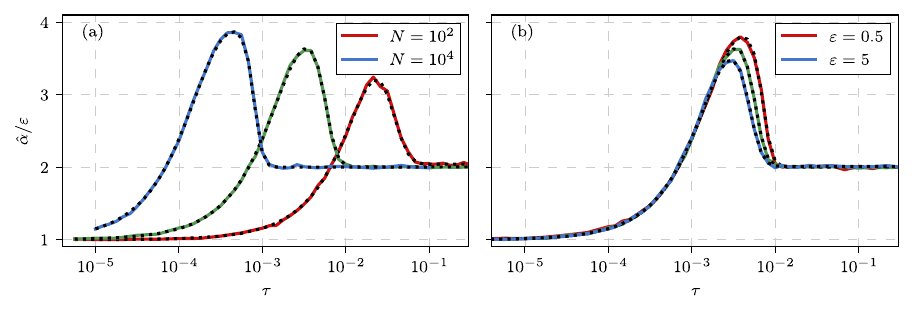}
\par\end{centering}
\caption{Scaling behavior of the normalized shape parameter estimate with respect
to the polling period: when the number of agents vary (a) and when
the independent transition rates vary (b). Solid colored curves correspond
to the results obtained by numerical simulation. Black dotted curves
indicate expected scaling behavior, given by Eq.~(\ref{eq:scaling-law}).
Base case simulation parameters (corresponds to the green curves in
both (a) and (b)): $\varepsilon_{0}=\varepsilon_{1}=\varepsilon=2$,
$N=10^{3}$. Legends indicate which parameter values differ from the
base case if any.\protect\label{fig:alpha-beta}}
\end{figure}

Solving
\begin{equation}
\mathrm{V}\left(\tau\right)=\mathrm{Var}\left[\mathrm{BetaBin}\left\{ N,\varepsilon_{1}L\left(\tau\right),\varepsilon_{0}L\left(\tau\right)\right\} \right],
\end{equation}
where $\mathrm{V}\left(\tau\right)=\mathrm{Var}\left[A_{\infty}\right]$
is introduced just as a notational convenience chosen to emphasize
the dependence of the stationary variance on $\tau$, with respect
to the scaling law yields
\begin{equation}
L\left(\tau\right)=\frac{\varepsilon_{0}\varepsilon_{1}N^{2}-\left(\varepsilon_{0}+\varepsilon_{1}\right)^{2}\mathrm{V}\left(\tau\right)}{\left(\varepsilon_{0}+\varepsilon_{1}\right)^{3}\mathrm{V}\left(\tau\right)-\varepsilon_{0}\varepsilon_{1}\left(\varepsilon_{0}+\varepsilon_{1}\right)N}.\label{eq:scaling-law}
\end{equation}
As can be seen in Fig.~\ref{fig:alpha-beta}, numerically simulated
scaling behavior of $\hat{\alpha}/\varepsilon$ coincides rather well
with the scaling law derived above.

\begin{figure}[tph]
\begin{centering}
\includegraphics[width=0.5\textwidth]{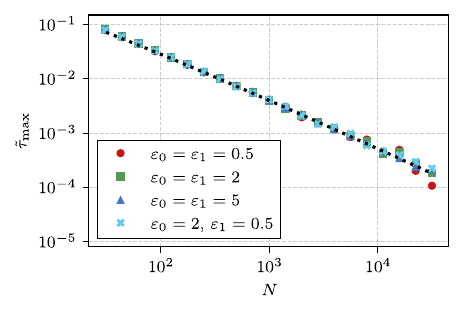}
\par\end{centering}
\caption{Normalized location of the peak in the numerically simulated $\hat{\alpha}\left(\tau\right)$
for different sets of $\varepsilon_{i}$ and varying $N$. Outcomes
of numerical simulations are depicted by varied symbols (corresponding
$\varepsilon_{i}$ values are given in the legend). The black dotted
curve corresponds to an appropriately normalized Eq.~(\ref{eq:tau-max}).\protect\label{fig:tau-max}}
\end{figure}

As can be seen in Figs.~\ref{fig:alpha-beta-epsi-asym}~and~\ref{fig:alpha-beta},
instead of observing a trivial sigmoid-like interpolation between
the small $\tau$ limit and the large $\tau$ limit, we observe a
non-trivial peak for some intermediate polling period $\tau_{\mathrm{max}}$.
As a comparison against the model with no announcement delay shows
(see Appendix~\ref{sec:no-delay-model}), the peak is induced by
the delay mechanism, while the doubling effect observed in the large
$\tau$ limit arises from the periodicity of the polls. The location
of this peak can be estimated by solving maximization problem on the
scaling law, Eq.~(\ref{eq:scaling-law}), or by solving minimization
problem on the stationary variance, Eq.~(\ref{eq:variance-ar2-stationary}).
Either approach produces the same rather complicated form, but as
usually we are interested in $N\gg\left(\varepsilon_{0}+\varepsilon_{1}\right)$
case, then a reasonably compact approximation can be employed
\begin{equation}
\tau_{\mathrm{max}}\approx\frac{1}{2\left(\varepsilon_{0}+\varepsilon_{1}+N\right)}\ln\left(\frac{3N}{\varepsilon_{0}+\varepsilon_{1}}\right).\label{eq:tau-max}
\end{equation}
In Fig.~\ref{fig:tau-max} we see that this approximation predicts
the location of the peak of numerically simulated $\hat{\alpha}/\varepsilon_{1}$
rather well. Note that in the figure we have normalized the location
of the peak to remove any dependency on $\varepsilon_{i}$, as only
then the results of different numerical simulations would be expected
to fall one a single curve. The normalization was done as follows:
$\tilde{\tau}_{\mathrm{max}}=\frac{\left(\varepsilon_{0}+\varepsilon_{1}+N\right)}{N}\cdot\tau_{\mathrm{max}}+\frac{\ln\left(\varepsilon_{0}+\varepsilon_{1}\right)}{2N}$.

From Eq.~(\ref{eq:scaling-law}), we can also deduce the maximum
value of the scaling law
\begin{equation}
L\left(\tau_{\mathrm{max}}\right)\approx4-\frac{6\left(1+\varepsilon_{0}+\varepsilon_{1}\right)}{1+3\left(\varepsilon_{0}+\varepsilon_{1}\right)+\sqrt{3N}}.
\end{equation}

As can be seen scaling law can have a maximum of at most $4$. The
maximum value is observed in the limits of $\varepsilon_{i}\rightarrow0$
or $N\rightarrow\infty$. If instead $\varepsilon_{i}\gg\sqrt{N}$,
the peak should notably diminish, but not disappear, as $L_{\mathrm{max}}$
would be slightly above $2$.

Let us also estimate the interval of polling periods for which the
peak is observed. Interval bounds could be determined by formulating
the problem with respect to the scaling law, but this would not yield
any intuition on the nature of the peak. Therefore, let us approach
this intuitively by exploring the behavior of expected value after
a single polling period of an arbitrary length. From Eq.~(\ref{eq:main-macro-rel})
with $k\tau=t$ and $s=\tau$, it is trivial to show that
\begin{equation}
\left\langle X\left(t+\tau\right)\right\rangle =\left\langle X\left(\infty\right)\right\rangle +\frac{\varepsilon_{0}^{\left(k\right)}X\left(t\right)-\varepsilon_{1}^{\left(k\right)}\left[N-X\left(t\right)\right]}{\varepsilon_{0}+\varepsilon_{1}+N}\exp\left[-\left(\varepsilon_{0}+\varepsilon_{1}+N\right)\tau\right].
\end{equation}
In the above $\left\langle X\left(\infty\right)\right\rangle $ stands
for the stationary expected value, which is reached in the $\tau\rightarrow\infty$
limit. Let us find such $\tau_{c}$ for which $\left\langle X\left(t+\tau_{c}\right)\right\rangle $
reaches (i.e., is almost indistinguishable from) the stationary expected
value:
\begin{equation}
\left\langle X\left(t+\tau_{c}\right)\right\rangle =\left\langle X\left(\infty\right)\right\rangle \pm\frac{1}{2}.
\end{equation}
Solving the above for $\tau_{c}$ yields:
\begin{equation}
\tau_{c}=\frac{\ln\left(2\left|\left\langle X\left(\infty\right)\right\rangle -X\left(t\right)\right|\right)}{\varepsilon_{0}+\varepsilon_{1}+N}.
\end{equation}
If we consider the largest possible distance between the initial and
stationary states, $\left|\left\langle X\left(\infty\right)\right\rangle -X\left(t\right)\right|=N$,
and the smallest $\left|\left\langle X\left(\infty\right)\right\rangle -X\left(t\right)\right|=1$,
we obtain:
\begin{equation}
\tau_{c}^{\left(1\right)}=\frac{\ln\left(2N\right)}{\varepsilon_{0}+\varepsilon_{1}+N},\quad\text{and}\quad\tau_{c}^{\left(2\right)}=\frac{\ln\left(2\right)}{\varepsilon_{0}+\varepsilon_{1}+N}.
\end{equation}
To verify whether these critical polling periods do indeed serve well
as the interval bounds, let us put them into the scaling law, Eq.~(\ref{eq:scaling-law}).
We obtain:
\begin{align}
L\left(\tau_{c}^{\left(1\right)}\right) & \approx2+\frac{2}{1+2\left(\varepsilon_{0}+\varepsilon_{1}\right)},\quad\text{and}\quad L\left(\tau_{c}^{\left(2\right)}\right)\approx2+\frac{\varepsilon_{0}+\varepsilon_{1}}{3N}.
\end{align}
It seems that $\tau_{c}^{\left(2\right)}$ is a really good estimate
for lower interval bound, and the result improves as $N$ grows larger.
$\tau_{c}^{\left(1\right)}$ is a worse estimate, but may serve as
a good rule of thumb for the determining the location of the upper
bound. This analysis seems to suggest that the overshoot in $L\left(\tau\right)$,
or alternatively the non-trivial decrease in $\mathrm{V}\left(\tau\right)$,
is caused by the tugging behavior between $A_{k-1}$ and $A_{k}$
becoming observable not only in individual trajectories, but for all
trajectories on average. This tugging behavior is absent from the
model with no delays (see Appendix~\ref{sec:no-delay-model}), while
in the model with delays it does diminish as the model starts to exhibit
behavior commonly observed in the large $\tau$ limit (i.e., effective
independence between even and odd polls).

Expanding on a parallel line of reasoning, one may introduce additional
$\tau_{c}$ by investigating instances when $\left\langle X\left(t+\tau_{c}\right)\right\rangle $
remains indistinguishable from $X\left(t\right)$. However, these
additional $\tau_{c}$ do not provide any new information. $\tau_{c}$
corresponding to the smallest distance between initial condition and
stationary value coincides with $\tau_{c}^{\left(2\right)}$, while
$\tau_{c}$ corresponding to the largest distance coincides with we
have discussed as a cutoff for the small $\tau$ limit.

\section{Periodic fluctuations induced by the polling mechanism\protect\label{sec:periodicity}}

Already in Fig.~\ref{fig:explain-polls} we have observed hints of
periodic fluctuations emerging from the model with delayed interactions.
In Section~\ref{sec:stationary-distributions} we have discussed
that the model with delayed interactions effectively becomes a second-order
Markov chain, which also suggests that the model could exhibit periodicity.
Furthermore, delays themselves may also be the cause of periodic fluctuations.

\begin{figure}[ph]
\begin{centering}
\includegraphics[width=0.9\textwidth]{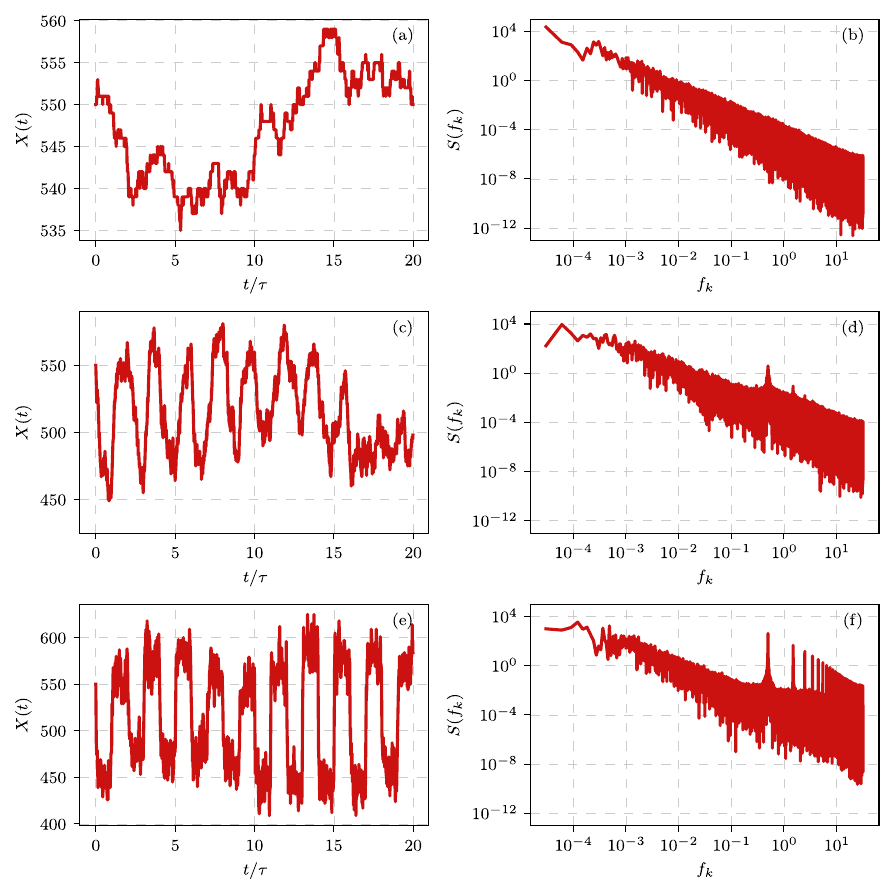}
\par\end{centering}
\caption{Fragment of time series obtained with different polling periods $\tau$
((a), (c) and (e)), and the respective power spectral density in the
poll index space ((b), (d) and (f)). Simulation parameters: $\varepsilon_{0}=\varepsilon_{1}=2$,
$N=10^{3}$, $A_{-1}=450$, $A_{0}=X\left(0\right)=550$ (in all cases),
and $\tau=3\cdot10^{-5}$ ((a) and (b)), $3\cdot10^{-3}$ ((c) and
(d)), and $3\cdot10^{-2}$ ((e) and (f)).\protect\label{fig:periodicity-sample}}
\end{figure}

As can be seen from a few sample trajectories shown in the plots on
the left side of Fig.~\ref{fig:periodicity-sample}, the larger $\tau$
the more immediately evident the periodic fluctuations in $X\left(t\right)$
become. Power spectral densities of these time series (the plots on
the right side of Fig.~\ref{fig:periodicity-sample}) indicate that
the main fluctuation frequency is $f_{k}=\frac{1}{2}$ in the poll
index space (hence, the subscript $k$), or $f=\frac{1}{2\tau}$ in
the physical time space. Additional peaks are observed at harmonic
frequencies of this main fluctuation frequency, therefore the only
source of periodicity is the periodic polling mechanism itself.

The emergence of the periodic fluctuations therefore can be quantified
by measuring the power spectral density at $f_{k}=\frac{1}{2}$. Which
is a squared absolute value of the Fourier transform at $f_{k}=\frac{1}{2}$:
\begin{equation}
S\left(f_{k}=\frac{1}{2}\right)=\frac{2}{Mf_{k}^{\left(s\right)}}\left|\sum_{m=0}^{M-1}\tilde{X}\left(\frac{m}{f_{k}^{\left(s\right)}}\tau\right)\cdot\exp\left[-2i\pi\frac{f_{k}}{f_{k}^{\left(s\right)}}m\right]\right|^{2}.
\end{equation}
In the above $M$ is the length of the time series, $f_{k}^{\left(s\right)}$
is the sampling frequency in the poll index space (corresponds to
the number of samples taken during a single polling period, e.g.,
$f_{k}^{\left(s\right)}=100$ in Fig.~\ref{fig:periodicity-sample}),
and $\tilde{X}\left(t\right)=\frac{X\left(t\right)-\left\langle X\right\rangle }{\sqrt{\mathrm{Var}\left[X\right]}}$,
i.e., standardized $X\left(t\right)$.

\begin{figure}[ph]
\begin{centering}
\includegraphics[width=0.5\textwidth]{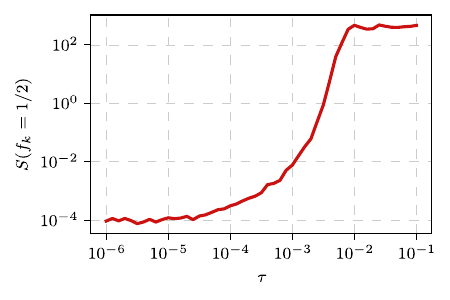}
\par\end{centering}
\caption{Power spectral density at the frequency corresponding to the observed
periodic fluctuations, $S\left(f_{k}=\frac{1}{2}\right)$, of numerically
simulated time series with different polling periods $\tau$. Simulation
parameters: $\varepsilon_{0}=\varepsilon_{1}=2$, $N=10^{3}$.\protect\label{fig:periodicity-psd}}
\end{figure}

The scaling of $S\left(f_{k}=\frac{1}{2}\right)$ concerning $\tau$
has a trivial sigmoid shape (see Fig.~\ref{fig:periodicity-psd}).
For small $\tau$, i.e., $\tau\lesssim\frac{2}{N\left(\varepsilon_{0}+\varepsilon_{1}+N\right)}$,
the system does not manage to incorporate the information potentially
revealed by polls. Therefore, no opinion swings are observed. Small
increases in $\tau$ do little to help the system incorporate the
polling information. Therefore, in this range, we observe a slow growth
of $S\left(f_{k}=\frac{1}{2}\right)$. For sufficiently large $\tau$,
i.e., $\tau\gg\tau_{c}^{\left(1\right)}$, further increases in $\tau$
have no effect observable effect, because the system already has had
enough time to incorporate all of the available polling information.
For intermediate $\tau$, even a small increase in $\tau$ allows
for more polling information to be incorporated, but the incorporation
is never complete. Consequently, in this range $S\left(f_{k}=\frac{1}{2}\right)$
grows rapidly, likely exponentially due to the exponential dependence
on $\tau$ in Eq.~(\ref{eq:single-agent-prob}). No special scaling
behavior is observed close to $\tau_{c}^{\left(2\right)}$, where
the peak in the scaling law starts.

\begin{figure}[ph]
\begin{centering}
\includegraphics[width=0.5\textwidth]{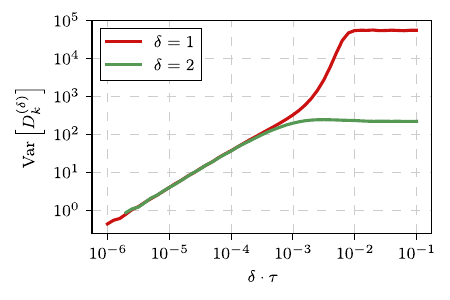}
\par\end{centering}
\caption{Variance of the differences between the consecutive poll outcomes,
$D_{k}^{\left(1\right)}$, and the differences between the next-consecutive
poll outcomes, $D_{k}^{\left(2\right)}$, with different polling periods
$\tau$. Simulation parameters: $\varepsilon_{0}=\varepsilon_{1}=2$,
$N=10^{3}$.\protect\label{fig:periodicity-diffs}}
\end{figure}

Alternatively, the periodic fluctuations could be quantified by looking
at the distribution of differences (or swings) between the consecutive
polls:
\begin{equation}
D_{k}^{\left(\delta\right)}=A_{k}-A_{k-\delta}.
\end{equation}
For the periodic fluctuations to become apparent the swings between
the consecutive poll outcomes, $D_{k}^{\left(1\right)}$, need to
become relatively large in comparison to the swings between the next-consecutive
poll outcomes, $D_{k}^{\left(2\right)}$. As can be seen in Fig.~\ref{fig:periodicity-diffs},
the variance of both $D_{k}^{\left(1\right)}$ and $D_{k}^{\left(2\right)}$
initially grows with $\tau$, but $D_{k}^{\left(2\right)}$ saturates
sooner, at approximately $\tau_{c}^{\left(2\right)}$. Saturation
of $D_{k}^{\left(1\right)}$ occurs for larger $\tau$, which roughly
corresponds to $\tau_{c}^{\left(1\right)}$. So difference in the
scaling behavior of the swing variances seems to be related to the
peak observed in $L\left(\tau\right)$.

\section{Conclusions\protect\label{sec:conclusions}}

We have examined the implications of information delay arising from
the periodic polling mechanism on the opinion formation process. We
have achieved this by integrating the periodic polling mechanism into
the noisy voter model. Specifically, we have replaced instantaneous
imitative interactions with imitative interactions mediated through
periodic polls. Consistent with real-world scenarios, we have assumed
a delay in announcing poll outcomes. Additionally, we have also aligned
the announcement delay with the polling period. The proposed generalization
constitutes a novel type of latency contrasting the one proposed in
\cite{Lambiotte2009PRE}. Namely, in \cite{Lambiotte2009PRE} it was
assumed that the agents freeze their opinion for some fixed duration,
while in the proposed generalization the perception of society as
a whole is being frozen for duration of the polling period instead.

The generalized model reveals non-trivial phenomenology. Namely, in
the short polling period limit, the generalized model is almost equivalent
to the noisy voter model. In the long polling period limit, the generalized
model retains approximately the Beta-binomial form of the stationary
distribution, but it becomes narrower (stationary variance is halved)
in comparison to what would be expected from the noisy voter model.
Yet the transition between the two limits exhibits non-monotonicity.
Namely, the stationary variance reaches its minimum, or the distribution
shape parameters reach their maximum, for some intermediate polling
period. These results were obtained both numerically and analytically.
It seems that the non-monotonic scaling behavior can be attributed
to the poll outcome announcement delay, as the model with no such
delay (see Appendix~\ref{sec:no-delay-model}) exhibits monotonic
scaling behavior. While the periodic poll outcome updates induce the
doubling effect observed in the large polling period limit.

For the purposes of numerical simulation, we have adapted the Gillespie
method, which is typically used for processes without delays \cite{Anderson2007JCP},
to incorporate delays specific to the generalized model. For short
polling periods, this method has served us well, but as the polling
periods become longer, the simulation using this method becomes increasingly
more time-consuming, because every transition needs to be explicitly
simulated. To address this issue, we have developed a macroscopic
simulation method, which allows us to simulate just the poll outcomes.
Furthermore, the proposed macroscopic simulation method has enabled
partial analytical treatment of the model. Python implementation of
all the algorithms used to simulate the generalized model (including
the semi-analytical approach) is available at \cite{DelayGithub}.

Given that delays themselves may contribute to periodic fluctuations,
we have investigated the emergence of periodicity within the generalized
model. We have found that the power spectral density peaks at the
frequency corresponding to the doubled polling period and at other
harmonic frequencies. The magnitude of the peak at the first harmonic
frequency scales predictably, following a monotonic sigmoid-like function
of the polling period. We have also examined the scaling of the variance
of the poll outcome swings. We have found that for short polling periods,
the variance of the consecutive and the next-consecutive poll swings
grows together, but the variance of the next-consecutive poll swings
saturates earlier and does so at the lower value. The polling period
at which the separation is observed roughly coincides with the start
of non-trivial behavior in the distribution shape parameter scaling
law. Saturation of the variance of the consecutive poll swings roughly
coincides with the long polling period limit.

The proposed generalization of the noisy voter model holds promise
for further refinement and a more comprehensive investigation of the
demonstrated rich phenomenology. The proposed generalization of the
noisy voter model could further serve as a foundational framework
for analytically probing other types of societal latency in the noisy
voter models. As mentioned, we explore a novel kind of latency, which
could be also introduced into other models of opinion dynamics, or
other generalizations of the voter model. Additionally, this extension
may prove instrumental in the development of domain-specific ARCH-like
models for opinion dynamics, branching away from the traditional applications
in economics and finance \cite{Giraitis2009,Chalissery2022JRFM,Bollerslev2023JEconom}.

{\small\begin{singlespace}
\section*{Author contributions}

\textbf{AK:} Conceptualization, Methodology, Software, Validation,
Writing, Visualization. \textbf{RA:} Methodology, Software, Validation,
Writing. \textbf{MR:} Conceptualization, Methodology, Validation,
Writing, Supervision. \textbf{FI:} Conceptualization, Writing, Supervision,
Project administration.


\appendix

\section{The large $\tau$ limit\protect\label{sec:large-tau-limit}}

Starting from the premise of the macroscopic simulation method introduced
in Section~\ref{subsec:macro-method}, we can explore the statistical
characteristics of the model in the limit of large $\tau$. Polling
period $\tau$ should be as large as it would make the outcome $A_{k+1}$
effectively independent of $A_{k}$. This can be achieved when $\tau\gg\frac{1}{\varepsilon_{0}+\varepsilon_{1}+N}$.
For $\tau\gg\frac{1}{\varepsilon_{0}+\varepsilon_{1}+N}$, we have
that $P_{1}\left(\tau|0\right)\approx P_{1}\left(\infty\right)$ and
$P_{1}\left(\tau|1\right)\approx P_{1}\left(\infty\right)$. As in
this limit the next poll outcome $A_{k+1}$ depends only on $A_{k-1}$,
we can now analyze the model with delayed interactions not as a second-order
Markov chain, but instead as two independent first-order Markov chains.
Because these two chains are identical in all regards except the initial
conditions, let us limit the detailed derivations to the chain for
even poll indices.

Let $p_{\mathcal{T}}\left(x|u\right)$ denote the probability of observing
$A_{k+1}=x$ given that $A_{k-1}=u$, i.e., the transition probability
between the outcomes of subsequent even or odd polls. The distribution
of the $k$-th poll outcome can be obtained through a recursive relationship
\begin{equation}
p_{k+1}\left(x\right)=\sum_{u=0}^{N}p_{\mathcal{T}}\left(x|u\right)p_{k-1}\left(u\right),\label{eq:dist-recursive}
\end{equation}
with the initial conditions
\begin{equation}
p_{0}\left(x\right)=\delta\left(x-A_{0}\right),\quad p_{-1}\left(x\right)=\delta\left(x-A_{-1}\right).
\end{equation}
In the expression above, $\delta\left(x\right)$ denotes a Kronecker
delta function, its value is $1$ if $x=0$ and is $0$ otherwise.
Then both $P_{1}\left(\tau|0\right)$ and $P_{1}\left(\tau|1\right)$
are close to $P_{1}\left(\infty\right)$, $p_{\mathcal{T}}\left(x|u\right)$
corresponds to a Binomial distribution probability mass function with
$N$ trials and the success probability $P_{1}\left(\infty\right)$,
which is implicitly a function of $u=A_{k-1}$, Eq.~(\ref{eq:single-agent-prob-stationary}).
Obtaining a general analytical expressions for $p_{i}\left(x\right)$
or $p_{\infty}\left(x\right)$ doesn't seem feasible, but this problem
could be approached from a numerical perspective. Approach discussed
in Section~\ref{subsec:semi-analytical-approach} would yield similar
results to iterating Eq.~(\ref{eq:dist-recursive}), but the approach
discussed in Section~\ref{subsec:semi-analytical-approach} would
not be limited to the large $\tau$ values. Instead let us focus on
deriving expressions for the evolution of mean and variance of the
poll outcome distribution.

As the poll outcome distributions are linked via recursive relationship,
Eq.~(\ref{eq:dist-recursive}), we can show that the mean will satisfy
another recursive relationship
\begin{equation}
\left\langle A_{k+1}\right\rangle =\sum_{x=0}^{N}xp_{k+1}\left(x\right)=\sum_{u=0}^{N}\left[\left(\sum_{x=0}^{N}xp_{\mathcal{T}}\left(x|u\right)\right)p_{k-1}\left(u\right)\right]=\sum_{u=0}^{N}\frac{N\left(\varepsilon_{1}+u\right)}{\varepsilon_{0}+\varepsilon_{1}+N}p_{k-1}\left(u\right)=\frac{N\left(\varepsilon_{1}+\left\langle A_{k-1}\right\rangle \right)}{\varepsilon_{0}+\varepsilon_{1}+N}.
\end{equation}
For even poll indices, this recursive form can be rewritten as
\begin{equation}
\left\langle A_{k}\right\rangle =\left\langle A_{\infty}\right\rangle +\left(A_{0}-\left\langle A_{\infty}\right\rangle \right)\left(\frac{N}{\varepsilon_{0}+\varepsilon_{1}+N}\right)^{k/2},\label{eq:big-tau-even-mean}
\end{equation}
where $k$ is the poll index, and $\left\langle A_{\infty}\right\rangle $
is the stationary mean,
\begin{equation}
\left\langle A_{\infty}\right\rangle =\frac{N\varepsilon_{1}}{\varepsilon_{0}+\varepsilon_{1}}.\label{eq:appendix-big-tau-stationary-mean}
\end{equation}
Repeating the same derivation for the odd poll indices, we obtain
\begin{equation}
\left\langle A_{k}\right\rangle =\left\langle A_{\infty}\right\rangle +\left(A_{-1}-\left\langle A_{\infty}\right\rangle \right)\left(\frac{N}{\varepsilon_{0}+\varepsilon_{1}+N}\right)^{\left(k+1\right)/2}.\label{eq:big-tau-odd-mean}
\end{equation}
In Fig\@.~\ref{fig:big-tau-dist}~(a) and (b) we see that the analytical
expressions for the mean of even and odd $k$ polls match the numerical
simulations rather well.

The recursive relationship for the second raw moment is somewhat more
complicated
\begin{align}
\left\langle A_{k+1}^{2}\right\rangle  & =\sum_{x=0}^{N}x^{2}p_{k+1}\left(x\right)=\sum_{u=0}^{N}\left[\left(\sum_{x=0}^{N}x^{2}p_{\mathcal{T}}\left(x|u\right)\right)p_{k-1}\left(u\right)\right]=\nonumber \\
 & =\frac{N\left[\varepsilon_{1}\left(\varepsilon_{0}+N+\varepsilon_{1}N\right)+\left(\varepsilon_{0}-\varepsilon_{1}+N+2\varepsilon_{1}N\right)\left\langle A_{k-1}\right\rangle +\left(N-1\right)\left\langle A_{k-1}^{2}\right\rangle \right]}{\left(\varepsilon_{0}+\varepsilon_{1}+N\right)^{2}}.
\end{align}
For even poll indices, this recursive form can be rewritten as
\begin{equation}
\left\langle A_{k}^{2}\right\rangle =A_{\infty}^{\left(2\right)}+\left[A_{0}^{2}-A_{\infty}^{\left(2\right)}-A_{\mathrm{mid}}^{\left(2\right)}\right]\left[\frac{N\left(N-1\right)}{\left(\varepsilon_{0}+\varepsilon_{1}+N\right)^{2}}\right]^{k/2}+A_{\mathrm{mid}}^{\left(2\right)}\left[\frac{N}{\varepsilon_{0}+\varepsilon_{1}+N}\right]^{k/2},
\end{equation}
with
\begin{align}
A_{\mathrm{mid}}^{\left(2\right)} & =\left(A_{0}-\left\langle A_{\infty}\right\rangle \right)\frac{\varepsilon_{0}-\varepsilon_{1}+N+2\varepsilon_{1}N}{\varepsilon_{0}+\varepsilon_{1}+1},\\
A_{\infty}^{\left(2\right)} & =\left\langle A_{\infty}^{2}\right\rangle =\frac{N\varepsilon_{1}\left[\left(\varepsilon_{0}+N+\varepsilon_{1}N\right)\left(\varepsilon_{0}+\varepsilon_{1}\right)+\left(\varepsilon_{0}-\varepsilon_{1}+N+2\varepsilon_{1}N\right)N\right]}{\left(\varepsilon_{0}+\varepsilon_{1}\right)\left[\left(\varepsilon_{0}+\varepsilon_{1}+N\right)^{2}-N\left(N-1\right)\right]}.
\end{align}
Given the expression for the second raw moment, the variance for the
even poll indices $k$ can be shown to be
\begin{align}
\mathrm{Var}\left[A_{k}\right]= & \left\langle A_{k}^{2}\right\rangle -\left\langle A_{k}\right\rangle ^{2}=\mathrm{Var}\left[A_{\infty}\right]+\left(A_{0}^{2}-A_{\infty}^{\left(2\right)}-A_{\mathrm{mid}}^{\left(2\right)}\right)\left[\frac{N\left(N-1\right)}{\left(\varepsilon_{0}+\varepsilon_{1}+N\right)^{2}}\right]^{k/2}+\nonumber \\
 & +\left[A_{\mathrm{mid}}^{\left(2\right)}-2\left\langle A_{\infty}\right\rangle \left(A_{0}-\left\langle A_{\infty}\right\rangle \right)\right]\left(\frac{N}{\varepsilon_{0}+\varepsilon_{1}+N}\right)^{k/2}-\left(A_{0}-\left\langle A_{\infty}\right\rangle \right)^{2}\left(\frac{N}{\varepsilon_{0}+\varepsilon_{1}+N}\right)^{2k}.\label{eq:big-tau-even-var}
\end{align}
Repeating the same derivation for the odd poll indices $k$, we see
that
\begin{align}
\mathrm{Var}\left[A_{k}\right]= & \mathrm{Var}\left[A_{\infty}\right]+\left(A_{-1}^{2}-A_{\infty}^{\left(2\right)}-A_{\mathrm{mid}}^{\left(2\right)}\right)\left[\frac{N\left(N-1\right)}{\left(\varepsilon_{0}+\varepsilon_{1}+N\right)^{2}}\right]^{\left(k+1\right)/2}+\nonumber \\
 & +\left[A_{\mathrm{mid}}^{\left(2\right)}-2\left\langle A_{\infty}\right\rangle \left(A_{-1}-\left\langle A_{\infty}\right\rangle \right)\right]\left(\frac{N}{\varepsilon_{0}+\varepsilon_{1}+N}\right)^{\left(k+1\right)/2}-\left(A_{-1}-\left\langle A_{\infty}\right\rangle \right)^{2}\left(\frac{N}{\varepsilon_{0}+\varepsilon_{1}+N}\right)^{2k+2}.
\end{align}
In the above expressions, for both even and odd poll indices $k$,
$\mathrm{Var}\left[A_{\infty}\right]$ stands for the stationary variance,
\begin{equation}
\mathrm{Var}\left[A_{\infty}\right]=\frac{N\varepsilon_{1}\varepsilon_{0}\left(\varepsilon_{0}+\varepsilon_{1}+N\right)^{2}}{\left(\varepsilon_{0}+\varepsilon_{1}\right)^{2}\left[\left(\varepsilon_{0}+\varepsilon_{1}\right)^{2}+\left(2\varepsilon_{0}+2\varepsilon_{1}+1\right)N\right]}.\label{eq:big-tau-stationary-var-exact}
\end{equation}
In Fig\@.~\ref{fig:big-tau-dist}~(c) we see that the analytical
expressions for the evolution of variance matches the numerical simulations
rather well.

While the stationary mean, Eq.~(\ref{eq:appendix-big-tau-stationary-mean}),
has a form identical to the mean of the Beta-binomial distribution,
the stationary variance has a bit different form. Yet if the number
of agents is large, $N\gg\left(\varepsilon_{0}+\varepsilon_{1}\right)$,
the difference in form becomes negligible. These results suggest that
$A_{\infty}$ could be well approximated by $\mathrm{BetaBin}\left(N,2\varepsilon_{1},2\varepsilon_{0}\right)$
distribution. This is confirmed by numerical simulation in Fig.~\ref{fig:big-tau-dist}~(d).

\section{Model with no poll outcome announcement delay\protect\label{sec:no-delay-model}}

The model presented in the main body of this paper assumes both periodic
polling and poll outcome announcement delay. For simplicity sake,
we have also assumed that the polling period and the announcement
delay are aligned and equal to $\tau$. This is a reasonable assumption
when the polls are conducted frequently, e.g., on daily or weekly
basis. If the polls are conducted not as frequently, then the announcement
delay could become negligible. In this appendix, let us briefly consider
the case when there is no announcement delay. In other words, let
the $k$-th poll be conducted and its outcome $A_{k}$ be announced
at $t=k\tau$. This assumption completely eliminates any explicit
dependence on the previous poll outcome $A_{k-1}$ present in the
main model.

If the poll outcomes are announced instantaneously (with no delay),
the system-wide transition rates take the following form
\begin{equation}
\lambda_{k}^{+}=\left(N-X\right)\left[\varepsilon_{1}+A_{k}\right],\qquad\lambda_{k}^{-}=X\left[\varepsilon_{0}+\left(N-A_{k}\right)\right].
\end{equation}
Likewise, $A_{k-1}$ is replaced by $A_{k}$ in the effective individual
agent transition rates, Eq.~(\ref{eq:individual-effective-rates}).
When considering a single polling period, as done in Section~\ref{subsec:macro-method},
the stationary probability of observing an agent in the ``$1$''
state would tend toward
\begin{equation}
P_{1}\left(\infty\right)=\frac{\varepsilon_{1}+A_{k}}{\varepsilon_{0}+\varepsilon_{1}+N}.
\end{equation}

With the explicit dependence on $A_{k-1}$ being eliminated, the model
can be approximated by a first-order auto-regressive process \cite{Brockwell1991Springer}.
From Eq.~(\ref{eq:main-macro-rel}), after taking into account the
differences between the two models, it can be shown that the conditional
mean of $A_{k+1}$ with respect to $A_{k}$ is given by
\begin{equation}
\left\langle A_{k+1}|A_{k}\right\rangle =\left[\varphi_{1}+\varphi_{2}\left(1-\varphi_{1}\right)\right]A_{k}+\varepsilon_{1}\varphi_{2}\left(1-\varphi_{1}\right).
\end{equation}
In the above $\varphi_{1}$ and $\varphi_{2}$ retain the same expressions
as discussed in the main body of the manuscript. Note that the expression
we have obtained has form which is mostly identical to Eq.~(\ref{eq:cond-mean})
(with $A_{k-1}$ being replaced by $A_{k}$). Therefore, averaging
the conditional mean over the stationary distribution, and then solving
for the stationary mean yields the same result as for the main model,
$\left\langle A_{\infty}\right\rangle =\frac{\varepsilon_{1}N}{\varepsilon_{0}+\varepsilon_{1}}$.

Following the same considerations as in Section~\ref{sec:stationary-distributions},
we see that the centered poll outcomes can be approximated by a first-order
auto-regressive process \cite{Brockwell1991Springer} of the following
form
\begin{equation}
\tilde{A}_{k+1}=\varphi_{3}\tilde{A}_{k}+\eta_{k+1}.
\end{equation}
In the above $\varphi_{3}=\varphi_{1}+\varphi_{2}\left(1-\varphi_{1}\right)$,
and $\eta_{k+1}$ is the error (white noise) term. From Yule-Walker
equations \cite{Brockwell1991Springer}, we have that the stationary
correlation between $\tilde{A}_{k+1}$ and $\tilde{A}_{k}$ has a
much simpler form than for the main model, i.e., $\rho_{1}=\varphi_{3}$.
Then the stationary variance of the poll outcome distribution is a
solution of
\begin{equation}
\mathrm{Var}\left[A_{\infty}\right]\left(1-\varphi_{3}^{2}\right)=\mathrm{Var}\left[\eta_{\infty}\right].\label{eq:appendix-variance-equation}
\end{equation}
From Eq.~(\ref{eq:main-macro-rel}), after taking into account the
differences between the two models, it can be shown that
\begin{equation}
\mathrm{Var}\left[\eta_{k}|A_{k}\right]=\psi_{0}+\left(\psi_{1}+\psi_{2}\right)\tilde{A}_{k}+\left(\psi_{12}+\psi_{22}\right)\tilde{A}_{k}^{2}.
\end{equation}
In the above $\psi_{0}$, $\psi_{1}$, $\psi_{2}$, $\psi_{12}$ and
$\psi_{22}$ retain the forms given in Eq.~(\ref{eq:psi-params}).
The obtained expression for the conditional variance is once again
similar to respective counterpart obtained for the main model Eq.~(\ref{eq:var-cond})
(with $\tilde{A}_{k-1}$ being replaced by $\tilde{A}_{k}$). Yet
averaging the above over stationary distribution gives a somewhat
different result:
\begin{equation}
\mathrm{Var}\left[\eta_{\infty}\right]=\psi_{0}+\left(\psi_{12}+\psi_{22}\right)\mathrm{Var}\left[A_{\infty}\right].
\end{equation}
Inserting the above into Eq.~(\ref{eq:appendix-variance-equation})
and solving for $\mathrm{Var}\left[A_{\infty}\right]$ yields
\begin{equation}
\mathrm{Var}\left[A_{\infty}\right]=\frac{\psi_{0}}{1-\psi_{12}-\psi_{22}-\left[\varphi_{1}+\varphi_{2}\left(1-\varphi_{1}\right)\right]^{2}}.\label{eq:appendix-instantenous-variance}
\end{equation}
As Eq.~(\ref{eq:scaling-law}) is written in a model independent
way, the scaling law can be obtained from it by replacing $V\left(\tau\right)$
with $\mathrm{Var}\left[A_{\infty}\right]$ from the above. As can
be seen in Fig.~\ref{fig:compare-laws}, the scaling law of the model
without the poll outcome announcement delay exhibits monotonic scaling
behavior. This allows us to conclude that non-monotonic scaling behavior
observed in the main model is induced by the announcement delay.

\begin{figure}[th]
\begin{centering}
\includegraphics[width=0.5\textwidth]{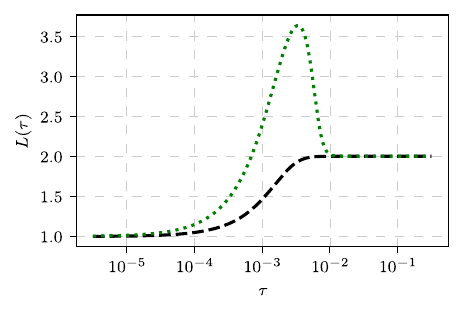}
\par\end{centering}
\caption{Scaling law obtained for the model with announcement delay (dotted
green curve) and for the model without delay (dashed black curve).
Scaling laws were obtained from Eq.~(\ref{eq:scaling-law}) by replacing
$V\left(\tau\right)$ with the model appropriate expression for the
stationary variance, Eq.~(\ref{eq:variance-ar2-stationary}) for
the model with delay and Eq.~(\ref{eq:appendix-instantenous-variance})
for the model without delay. Parameters correspond to the base case
considered in the main body of this paper (i.e., $\varepsilon_{0}=\varepsilon_{1}=\varepsilon=2$,
$N=10^{3}$).\protect\label{fig:compare-laws}}
\end{figure}

\end{singlespace}}

\end{document}